\documentclass[preprint,showpacs,preprintnumbers]{revtex4}
\usepackage{amssymb}
\usepackage{amsmath}
\usepackage{graphicx}
\usepackage{dcolumn}
\usepackage{bm}

\input{tcilatex}

\begin{document}

\title{ Experimental realization of a highly secure chaos communication
under strong channel noise}
\author{Weiping Ye$^{1}$, Qionglin Dai$^{1}$, Shihong Wang$^{2}$, Huaping L%
\"{u}$^{2}$, Jinyu Kuang$^{1}$, Zhenfeng Zhao$^{1},$ Xiangqing Zhu$^{1}$,
Guoning Tang$^{2}$, Ronghuai Huang$^{1}$, ang Gang Hu$^{2,3,\dagger }$}
\affiliation{$^{1}$Department of Electronics, Beijing Normal University, Beijing, 100875,
China\\
$^{2}$Department of Physics, Beijing Normal University, Beijing, 100875,China}
\affiliation{$^{3}$The Key Laboratory of Beam Technology and Material\\
$^{\dagger }$Correspondent author (email:ganghu@bnu.edu.cn)}
\date{\today }

\begin{abstract}
A one-way coupled spatiotemporally chaotic map lattice is used to contruct
cryptosystem. With the combinatorial applications of both chaotic
computations and conventional algebraic operations, our system has optimal
cryptographic properties much better than the separative applications of
known chaotic and conventional methods. We have realized experiments to
pratice duplex voice secure communications in realistic Wired Public
Switched Telephone Network by applying our chaotic system and the system of
Advanced Encryption Standard (AES), respectively, for cryptography. Our
system can work stably against strong channel noise when AES fails to work.

\end{abstract}

\maketitle

\section{\protect\bigskip Introduction}

Chaotic systems have several significant features favorable to secure
communications, such as aperiodicity (useful for one-time pad cipher);
sensitivity to initial condition and parameters (useful for effective bit
confusion and diffusion \cite{1}); and random-like behavior (useful for
producing output with satisfactory statistics). With all these advantages
scientists expected to introduce new and powerful tools of chaotic
cryptography \cite{2,3,4,5,6,7}. Nevertheless, during the last decade, many
pitfalls and drawbacks of cryptosystems based on chaos synchronization have
been found. The main problems are: low security due to easy reconstruction
of chaotic dynamics \cite{8,9,10,11,12}, slow performance speed due to
analytical floating-point computation, and weak resistance against channel
noise due to large bit error propagation caused by finite chaos
synchronization time. Recently, various methods have been suggested to solve
the above problem \cite{13,14,15,16,17}. In this paper we propose to use a
one-way coupled chaotic map system to construct a cryptosystem with optimal
overall properties. The crucial merits of this system are: on one hand we
use spatiotemporal chaos to fully apply and develop the advantages of
chaotic cryptography, and on the other hand we incorporate some simple
algebraic operations in the conventional cryptography to overcome the
disadvantages of analytical chaotic computations. With the combinative
applications of chaotic and conventional methods our system has optimal
cryptographic properties much better than the separative applications of
chaotic and conventional methods known so far. We design an experiment set
with embedded CPUs and use this set to practice duplex voice secure
communications in realistic wired Public Switched Telephone Network, by
applying our chaotic cryptosystem and the system of the Advanced Encryption
Standard (AES) \cite{18}, respectively, for comparisons. Experimental and
numerical results show that our system is considerably better than AES with
both security and performance speed. Most significantly, our system can work
stably against strong channel noise when AES fails to work.

\section{Spatiotemporal Chaotic Cryptosystem (STCC)}

We take a one-way coupled map lattice for spatiotemporal-chaos-based
cryptography, which has the encryption transformation as

\begin{eqnarray}
x_{n+1}(j) &=&(1-\varepsilon )f_{j}[x_{n}(j)]+\varepsilon f_{j}[x_{n}(j-1)],
\notag \\
f_{j}(x) &=&(3.75+a_{j}/4)x(1-x),\text{ }a_{j}\in \lbrack 0,1],  \notag \\
\text{ }j &=&1,\cdots ,m  \TCItag{1a} \\
x_{n}(0) &=&D_{n}/2^{9}+0.1  \notag
\end{eqnarray}

\begin{eqnarray}
x_{n+1}(m+1) &=&(1-\varepsilon )f[x_n(m+1)]+\varepsilon f[x_n(m)],  \notag \\
Q_n^{\prime } &=&[\text{int}(x_n(m+1)\times 2^{52})]\text{ \ }\func{mod}
2^{32}  \TCItag{1b} \\
Q_n &=&Sbox(Q_n^{\prime })  \notag \\
f(x) &=&4x(1-x),\text{ \ \ }z_n=Q_n/2^{32}  \notag
\end{eqnarray}

\begin{eqnarray}
x_{n+1}(m+2) &=&(1-\varepsilon )f_{1}[x_{n}(m+2)]+\varepsilon f_{1}(z_{n}), 
\notag \\
x_{n+1}(j) &=&(1-\varepsilon )f_{j-m-1}[x_{n}(j)]  \notag \\
&&+\varepsilon f_{j-m-1}[x_{n}(j-1)],  \TCItag{1c} \\
j &=&m+3,\cdots ,2m+1  \notag
\end{eqnarray}

\begin{eqnarray}
x_{n+1}(j) &=&(1-\varepsilon )f[x_n(j)]+\varepsilon f[x_n(j-1)],  \TCItag{1d}
\\
j &=&2m+2,\cdots ,2m+2N  \notag
\end{eqnarray}

\begin{eqnarray}
y_{n+1}(1,1) &=&(1-\varepsilon )f[y_{n}(1,1)]+\varepsilon f[x_{n}(2m+2N)] 
\notag \\
y_{n+1}(j_{1},1) &=&(1-\varepsilon )f[y_{n}(j_{1},1)]  \notag \\
&&+\frac{\varepsilon }{2}\left\{
f[y_{n}(j_{1}-1,1)]+f[x_{n}(2m+2j_{1}-1)]\right\}  \notag \\
y_{n+1}(1,j_{2}) &=&(1-\varepsilon )f[y_{n}(1,j_{2})]  \notag \\
&&+\frac{\varepsilon }{2}\left\{
f[y_{n}(1,j_{2}-1)]+f[x_{n}(2m+2j_{2}-2)]\right\}  \TCItag{1e} \\
y_{n+1}(j_{1},j_{2}) &=&(1-\varepsilon )f[y_{n}(j_{1},j_{2})]  \notag \\
&&+\frac{\varepsilon }{2}\left\{
f[y_{n}(j_{1},j_{2}-1)]+f[y_{n}(j_{1}-1,j_{2})]\right\}  \notag \\
\ \ \;\ \ j_{1},j_{2} &=&2,3,\cdots ,N  \notag
\end{eqnarray}

\begin{eqnarray}
\text{\ }K_{n}(j_{1},j_{2}) &=&\text{int}[y_{n}(j_{1},j_{2})\times 2^{52}]%
\text{ \ mod }2^{32},  \notag \\
S_{n} &=&[K_{n}(j_{1},j_{2})+I_{n}(j_{1},j_{2})]\text{ \ }\func{mod}2^{32}, 
\notag \\
\text{ }j_{1},j_{2} &=&1,2,\cdots ,N  \TCItag{1f} \\
D_{n} &=&[S_{n}(N,N)\gg 24]\&255  \notag
\end{eqnarray}%
where the S-box is defined as

\begin{eqnarray}
A_1 &=&[(Q_n^{\prime }\gg 24)\&255],A_2=[(Q_n^{\prime }\gg 16)\&255],  \notag
\\
A_3 &=&[(Q_n^{\prime }\gg 8)\&255],A_4=[(Q^{\prime }\&255],  \notag \\
A_0 &=&A_1\oplus A_2\oplus A_3\oplus A_4  \TCItag{2} \\
Q_n &=&[A_0\ll 24]+[A_4\ll 16]+[A_3\ll 8]+A_2  \notag
\end{eqnarray}
The operation $x\gg y$ ($x\ll y)$ denotes a right (left) shift of $x$ by $y$
bits, the \& operator is bitwise AND, and $\oplus $ means bitwise XOR.

The decryption system is driven by the transmitted signal as

\begin{equation*}
x_n^{\prime }(0)=D_n/2^9+0.1
\end{equation*}
and all other dynamic forms of the receiver are exactly the same as those of
the transmitter with $x_n(j),$ $z_n,$ $y_n(j_1,j_2),$ $K_n(j_1,j_2),$ $%
I_n(j_1,j_2),$ and $\mathbf{a=(}a_1,a_2,\cdots ,a_m)$ replaced by $%
x_n^{\prime }(j),$ $z_n^{\prime },$ $y_n^{\prime }(j_1,j_2),$ $K_n^{\prime
}(j_1,j_2),$ $I_n^{\prime }(j_1,j_2),$ and $\mathbf{b=(}b_1,b_2,\cdots
,b_m), $ respectively. With $\mathbf{b=a,}$ the receiver can reach chaos
synchronization with the transmitter, and successfully recover the true
plaintext as

\begin{eqnarray}
\mathbf{b} &\mathbf{=a,}&\text{ }y_{n}^{\prime
}(j_{1},j_{2})=y_{n}(j_{1},j_{2}),\text{ }  \notag \\
K_{n}^{\prime }(j_{1},j_{2}) &=&K_{n}(j_{1},j_{2}),\text{ }I_{n}^{\prime
}(j_{1},j_{2})=I_{n}(j_{1},j_{2})  \TCItag{3}
\end{eqnarray}%
In Eqs.(1)-(3) three parameters $\varepsilon ,$ $m,$ $N$ are adjustable for
controlling different cryptographic properties of the system, according to
the actual requirements of realistic secure communications. In this paper,
we fix

\begin{equation}
\varepsilon =0.99,\text{ \ \ }m=3,\text{ \ }N=4  \tag*{(4)}
\end{equation}
and will simply call our spatiotemporally chaotic cryptosystem Eq.(1) with
parameters (4) as STCC. The scheme of STCC is shown in Fig.1, and the
decryption system has exactly the same structure with feedback structure of
the transmitter replaced by driving structure in the receiver. The former is
thus a high-dimensional hyperchaos while the latter becomes nonchaotic with
all conditional Lyapunov exponents negative.

The important and new point of system (1) is that we apply both
floating-point analytical computation of spatiotemporal chaos and algebraic
operations of integer numbers to construct our cryptosystem which possess
the advantages of both chaotic and conventional cryptographies.

First, we use high-dimensional spatiotemporal chaos as the basic structure
of the cryptography, which leads to the following significant advantages.
(i) Due to the high-dimensionality and chaoticity, the output keystreams and
ciphertexts have high complexity, long periodicity of computer realization
of chaos, and effective bit confusion and diffusion in many directions in
the variable space. All these properties are favorable to achieve high
practical security \cite{17}. (ii) Due to the extended nature of STCC we are
able to use many sites ($N\times N=16$ square sites in Fig.1) to produce
keystreams in parallel and greatly increase the speed of performance \cite{6}%
. (iii) With one-way coupled maps and strong coupling ($1-\varepsilon \ll 1$%
), the receiver can easily reach chaos synchronization with the transmitter
by a single driving $D_{n}$. Note, for each iteration the driving $D_{n}$
has only 8 bits while the total ciphertext 512 bits. This separation of
driving bits from nondriving ciphertext bits makes the communication well
resistant against strong channel noise. This point will be the central focus
later in our experiment.

After the above advantages of STCC, the following algebraic operations of
Eq.(1) can further and greatly improve the cryptographic properties of the
system. (i) In Eq.(1f) we apply an algebraic operation \textit{int, }which%
\textit{\ } makes all keystreams $K_{n}(j_{1},j_{2})$, ciphertext $%
S_{n}(j_{1},j_{2})$, and driving signal $D_{n}$ integer numbers. These
integralizations are extremely important for the robustness of highly secure
communications against computer round-off errors and channel noise \cite{14}%
. (ii) In Eqs.(1b) and (1f) we apply \textit{modulo} operations \cite{13,16}%
, which can considerably enhance the key sensitivity of the system, and can
also effectively improve the random-like statistics of the transmitted
signals. (iii) In Eqs.(1b) and (2) we incorporate a S-box algebraic
operation \cite{17}, which makes any analytical solution aiming at exposing
the secret key extremely difficult. All the algebraic operations (i)-(iii)
have been popularly used in the conventional cryptography. These operations
are so simple that they need very low computational expenses; and so weak
that they cannot play significant role in the conventional cryptography by
themselves. However, incorporating with the analytical computerization of
STCC, these simple algebraic operations play important roles in optimizing
the cryptographic properties of the system, because they are just suited,
with very little cost, to overcome the weakness of chaotic cryptography
mentioned in the introduction and allow the advantages of STCC fully
developed.

\section{Cryptographic properties of STCC}

We have evaluated various cryptographic properties of system (1).
Specifically, we have analyzed in details its security, performance, and
robustness, and compared these properties with those of AES. It is found
that STCC is considerably better than AES in all the above essential aspects.

(A) Security

We have evaluated the security of STCC by trying various effective attacks
based on key-sensitivity analysis; statistical-property analysis; and
analytical-solution analysis with the conditions of public-structure and
known plaintext, and find that no any tested method can be more effective
than the brute force attack. The detail of these evaluations (in particular,
the atatistics-based evaluations) will appear elsewhere \cite{17}$.$ In this
paragraph we focus on the key sensitivity analysis by using the error
function attack \cite{16}.

Since we consider public-structure and plaintext-known attacks, any intruder
can run the receiver system with the test key $\mathbf{b}$ to produce $%
I_{n}^{\prime }(j_{1},j_{2}),$ and then compare the output $I_{n}^{\prime
}(j_{1},j_{2})$ with the true plaintext $I_{n}(j_{1},j_{2})$ for exposing
the location of $\mathbf{a}$. Specifically, the intruder can compute the
following error function

\begin{equation}
e(j_{1},j_{2};\mathbf{b})=\frac{1}{T}\sum_{n=1}^{T}\left| i_{n}^{\prime
}(j_{1},j_{2})-i_{n}(j_{1},j_{2})\right| \text{ }  \tag{5}
\end{equation}

\begin{equation*}
\text{\ }i_{n}(j_{1},j_{2})=\frac{I_{n}(j_{1},j_{2})}{2^{32}},\text{ \ }%
i_{n}^{\prime }(j_{1},j_{2})=\frac{I_{n}^{\prime }(j_{1},j_{2})}{2^{32}}
\end{equation*}%
The secret key $\mathbf{a}$ can be extracted by minimizing the error
function as

\begin{equation}
e(j_{1},j_{2};\mathbf{b})=0\text{ \ \ \ \ at \ \ }\mathbf{b=a}  \tag{6}
\end{equation}%
This evaluation is called as the error function attack (EFA), which can be
used to analyze the key sensitivity property of the system.

In Figs.2(a) and (b) we fix $b_{2}=b_{3}=a_{1}=a_{2}=a_{3}=0.5$ and plot $%
e(1,1;b_{1})$ vs $b_{1}$ with $T=10^{8}$ for different detect resolutions.
It is clearly shown that $e(1,1;b_{1})$ raises rapidly to $\frac{1}{3}$ with
very small fluctuation for extremely small mismatch $\left|
b_{1}-a_{1}\right| \geq 2^{-45}.$ The same behavior can be observed as well
for $b_{2}$ and $b_{3.}$ In Fig.(2c) we fix $b_{3}=a_{3}=0.5$ and plot $%
e(1,1;b_{1},b_{2})$ vs $b_{1}$ and $b_{2}$, and observe a needle-like basin
exactly at $b_{1}=a_{1},b_{2}=a_{2}.$ In Fig.2(d) we present the behavior of 
$e(1,1;\mathbf{b})$ in the 3D parameter space. It is shown again that
whenever $\left| \mathbf{b}-\mathbf{a}\right| \geq 2^{-45}$ in the 3D space,
the error function raises immediately to about $\frac{1}{3}.$ Therefore, the
effective key number of our system against EFA is (2$^{45})^{3}=2^{135}$. It
can be easily proven that two data sequences, completely uncorrelated and
purely random and uniformly distributed in [0,1], have error value of Eq.(5)
equal to $\frac{1}{3}.$ The behavior of Figs.2(a)-(d) show convincingly
excellent key-parameter-sensitivity and satisfactory random-like statistical
properties. The cost for the intruder to break the security of our system by
using EFA is quantitated as

\begin{equation}
Cost=2^{135}\approx 10^{40}  \tag{7}
\end{equation}%
which is also the cost of the brute force attack for the 2$^{135}$ key
number.

Chaotic system (1) has a significant advantages over AES with security. The
security level of system (1) can be conveniently and greatly increased.
Simply increasing $m$ in Eq.(4) by one, we can surely enlarge the key number
(i.e., the level of security) by $2^{45}$ times, with the cryptographic
structure of Eq.(1) kept unchanging and with almost no increase (about $5\%$
increase) of computational cost. Thus, the security of STCC is practically
unshakable by the quick technology advance of attack machines, including
possible future quantum computers. In comparison, in order to greatly
increase the security level of AES, some other cryptographic properties have
to be sacrificed in balance.

(B) Encryption (decryption) speed

Usually, the floating-point analytical computation of real variables used in
chaotic cryptography is considerably slower than the algebraic operations of
integer numbers used in conventional cryptography. The encryption speed of
the former is thus often not comparable with that of the latter when the
securities of both systems are comparable. Nevertheless, STCC has rather
fast speed, because it fully takes the advantages of spatiotemporal chaos in
performance. By keeping high security, STCC produces ciphers in every
iteration (one-round encryption structure), and meanwhile in each iteration
many [$4\times 4=16$ for Eq.(4)] sites make encryption operations in
parallel. Therefore, with software implementation our STCC has very high
speed, higher than AES (which, with key of 128 bits, takes 10 rounds for
producing ciphers of a block). Specifically our STCC can encrypt 914Mbit and
430Mbit per second with 2GHz (A) and 700MHz (B) CPU computers, respectively,
while AES (with 128-bit key length and 128-bit block length) produces
267Mbit and 96Mbit ciphers for the same computers. STCC is therefore faster
than AES for 3.4 and 4.4 times with computers A and B, respectively.

A crucial point for the validity of the parallel encryption operations in
Fig.1 is that all the keystreams produced by the 16 square sites should be
practically uncorrelated from each other. We checked this point and found
that these keystreams are uncorrelated from each other and insuppressible
indeed, and this validates the parallel encryptions of Eq.(1) and Fig.1.

(C) Robustness of communications against channel noise

With the extremely high sensitivity shown in Fig.2, the problem of
robustness and reliability of secure communication against computer
round-off errors and channel noise should be carefully examined. It is well
known that all block-cipher systems and stream-cipher systems with
self-synchronizing scheme have a problem of bit error propagation (or say,
bit error avalanche), i.e., one bit error in the driving signal may cause a
large number of error bits in the received plaintext. In this regard, STCC
has some essential advantages. The most significant feature of our system is
that among the ciphers of 512 bits produced in each iteration [$%
S_{n}(j_{1},j_{2}),j_{1},j_{2}=1,2,3,4$ ] only 8 bits ($D_{n}$) are used for
driving. Therefore, only $\frac{1}{64}$ transmitted bits (driving bits) have
bit error avalanche problem, and all other bits (nondriving ciphertext bits)
have not. Hence, in average the avalanche destruction can be considerably
reduced in our case. In order to reduce the avalanche effect people must
include some additional bits for protection of the driving signal, and this
increases the cost of both cryptography and signal transmission. In doing so
our STCC has a great advantage over AES because in AES one should protect
all transmitted bits (each error bit of the ciphertext has an equal error
avalanche of 128 bits in the receiver plaintext) while for our system only
the driving bits, i.e., $8$ driving bits among the total $512$ cipher bits,
have the avalanche effect and need to be particularly protected. This
advantage will be shown, in our following experiment, to be extremely
important for the secure communications under strong channel noise.

\section{Experiments and Comparisons of STCC and AES}

Now we come to the central part of the present paper: the experimental
realization of STCC and the experimental comparisons between STCC and AES
for robustness against channel noise. We have realized a duplex voice
communication by using the Public Switched Telephone Network wired(PSTN).
The scheme of the experimental set is presented in Fig.3, where the
following significant points should be emphasized.

(i) In Fig.3 we use embedded CPUs connecting to other communication tools.
These CPUs perform cryptographic operations as well as other communication
tasks. Since the embedded CPU technique has been widely used in practical
communications, the experimental set of Fig.3 is realistic for applications.

(ii) We use the realistic PSTN for practicing secure communications.
Moreover, we intentionally add strong noise into the transmission channel to
study the possibility of secure communications in wireless telephone systems
where the channel noise is usually much stronger than the wired ones.

(iii) For the cryptographic part of the experimental secure communications,
we apply both STCC and AES, respectively, for comparisons. In order to
strengthen the resistance of the communications against channel noise, we
add some additional bits in Channel Coding, for protecting the driving
signals of both STCC and AES systems.

All the above arrangements are closely related to practice realistic secure
communication service.

For the channel environment we assume additive white Gaussion noise, which
yields certain fixed bit error rate (BER) of each transmitted bit.
Therefore, we will directly vary error probability $p$ of the transmitted
signal bits to model the noise perturbation in the channel. Moreover, for
the voice transmission from User A to User B we introduce BER before the
part of \textquotedblright Modulation\textquotedblright\ of User A in Fig.3
rather than after \textquotedblright Modulation\textquotedblright\ in the
channel, for the sake of convenience of experimental performances and
measurements. This arrangement does not change any essence since we are
interested only on the influence of different cryptographies not Modulation
and Demodulation operations.

In the part of Channel Coding we apply the standard approach of bit-error
correction \cite{19}. In case of AES, for transmitting a block of $128$
cipher bits we actually transmit $136$ total bits, of which $8$ additional
noncipher bits are used for correcting one bit error among the all 136
transmitted bits$^{19}$. The efficiency of the signal transmission is thus
reduced by $6\%$ (i.e., $6\%$ transmitted bits do not contain plaintext
information). This bit protection fails when two or more than two error bits
appear in a single block of ciphertext. In case of STCC, we protect the
driving bits $D_{n}$ only, and leave other cipher bits $S_{n}(j_{1},j_{2})$
unprotected. For transmitting $512$ ciphertext bits in each iteration, we
add $20$ additional bits to protect the $8$ driving bits. This protection
can correct maximum $5$ error bits in $28$ bits$^{19}$. With this driving
bit protection the efficiency of the signal transmission is reduced by about 
$4\%$.

Before the experiment of secure communication, we first examine the
behaviors of normal nonsecure (without cryptography) communication (NNC)
with various BER $p$'s. The working qualities of NNC for different ranges of
noise can be ranked subjectively and roughly by ears as: excellent for \ $%
p\lesssim \frac{1}{250}$; fairly well for $\frac{1}{250}<p\lesssim \frac{1}{%
100}$; bad for $\frac{1}{100}<p\lesssim \frac{1}{30}$; complete failure of
voice communication for $\ p>\frac{1}{30}$. Therefore, we will compare the
results of cryptographies of STCC and AES, in the range of $p$, $\frac{1}{%
2000}\leq p\leq \frac{1}{10}$.

In Fig.4(a) we plot $p_{N}$, $p_{S}$ and $p_{A}$ vs BER probability $p$
without bit protection, where $p_{N}$, $p_{S}$ and $p_{A}$ are the bit error
rates of NNC, STCC and AES secure communications, respectively. We have $%
p_{N}\approx p$ for all $p$ values. This is reasonable since the
\textquotedblright Modulation\textquotedblright\ and \textquotedblright
Demodulation\textquotedblright\ functions in Fig.3 do not observably change
the channel BER. Both $p_{S}$ and $p_{A}$ are much larger than $p_{N}$ due
to the error propagation effects (note, $p_{S,A}\approx 0.5$
implies/complete loss of the transmitted information). It is observed that $%
p_{S}$ and $p_{A}$ are in the same order. Without bit protection $p_{S}$ is
slightly larger than $p_{A},$ indicating that STCC has larger bit error
propagation rate than that of AES.

In Fig.4(b) we do the same as Fig.4(a) except that $p_{S}$ and $p_{A}$ are
measured with the function of bit error correction operating in Channel
Coding and Decoding parts. With the designed bit protections both $p_{S}$
and $p_{A}$ become considerably smaller than those in Fig.4(a). For STCC it
is striking that the bit errors of the received plaintext are reduced so
much that $p_{S}$ is almost identical to $p_{N}$ for $p\lesssim \frac{1}{30}$%
. This indicates that with the driving bit protection, the secure
communication based on STCC can work as good as NNC without suffering from
the bit error avalanche effect, whenever the normal communication works. To
our knowledge, it is the first time that a highly secure system has such
strong resistance against channel noise. In comparison, AES has much weaker
resistance against channel noise. $p_{A}$ is considerably larger than $p_{S}$
for all range of $p\geq \frac{1}{2000}$. From the figure we anticipate that
with AES secure communications with the designed bit error correction
function fail at $p$ of order $10^{-2}$, at which NNC and STCC may still
work. This distinction is significant in practice because wireless
communications may encounter channel noise close to this range.

As an example we transmit a female voice \textquotedblright
welcome\textquotedblright\ by applying the experiment set of Fig.3. The
input signal shown in Fig.5(a) is measured at gate G3 of Fig.3, while the
output signals are measured at gate G4 of Fig.3. Figures 5(b) and (c) show
the results without cryptography and with channel bit error rates $p=\frac{1%
}{100}$ and $\frac{1}{30}$, respectively. The characteristic features of the
input are kept in the output even as BER is up to $p=\frac{1}{30}$. In
Figs.5(d) and (e) we do the same as (b) and (c), respectively, by including
STCC cryptography and the error correction of driving bits (20 additional
bits for 512 ciphertext bits). There are almost no observable deviations
between STCC and NNC for both $p$'s. In Fig.5(f) and (g) we do the same as
(b) and (c), respectively, by including AES cryptography and the
corresponding transmitted bit protection (8 additional bits for a block of
128 ciphertext bits). In sharp contrast, the characteristics of the input
signal are essentially lost in (f) at $p=\frac{1}{100}$ and completely lost
in (g) at $p=\frac{1}{30}$. From the experimental results of Figs.4 and 5 it
is concluded that STCC can work much better than AES under strong channel
noise.

If we set an extremely small mismatch of the encryption and decryption keys
for STCC, e.g., $b_{1}=a_{1}+2^{-52}$, we observe complete loss of the
transmitted information (pure noise) even if the channel noise is zero ($p=0$
). This confirms the high key-sensitivity as well as high security of STCC.
This sensitivity is also observed for AES experiment.

\section{Conclusions}

In conclusion we have suggested a cryptosystem which basically uses
analytical floating-point computation and auxiliarily incorporates some
algebraic operations into the basic chaotic dynamics. These combinative
applications of chaotic and conventional cryptographic methods fully develop
the advantages of the chaotic crytography and overcome its disadvantages,
and thus achieve optimal overall cryptographic properties of high security,
fast performance speed, and strong resistance against channel noise and
other instabilities, which are considerably better than separative
applications of both chaotic and conventional cryptosystems known so far,
including AES. We have carried out an experiment practicing duplex voice
secure communications in the wired PSTN. By intentionally increasing channel
noise we have examined the possibility of highly secure communications in
the environment of strong channel noise. It is experimentally confirmed that
our chaotic cryptosystem works satisfactorily whenever normal nonsecure
communication successfully works. In a range of strong channel noise ($%
p\approx 10^{-2}$) assumed to be encountered by some wireless
communications, our system can satisfactorily perform the tasks of secure
communications while AES fails to work for the same bit transmission
efficiency.

This work was supported by the National Natural Science Foundation of China
under 10175010 and by Nonlinear Science Project.

\strut

\newpage\ \ \ \ \ \ \ \ \ \ \ \ \ \ \ \ \ \ \ \ \ \ \ \ \ \ \ \ \ \ \
Captions of Figures

Fig.1. Scheme of STCC encryption. The system is constructed with a 1D chain
of length 14 and a 2D $4\times 4$ network. In the 1D chain the six empty
triangles ($\Delta $) represent maps with the key parameters $a_{1},$ $%
a_{2}, $ and $a_{3}$; the black triangle ($\blacktriangle $) performs modulo
and S-box operations Eq.(1b); and the seven empty circles ($\bigcirc $) are
used for coupling the side sites of the 2D network. All the square sites ($%
\square $) in the 2D network perform encryption operations Eq.(1f)
simultaneously, among which the site (4,4) (black square $\blacksquare $)
produces driving signal $D_{n}$ according to Eqs.(1f) and (1a). All the
solid arrows ($\rightarrow $) denote coupling directions; $K,$ $I,$ and $S$
indicate the keystream, plaintext, and ciphertext, respectively.

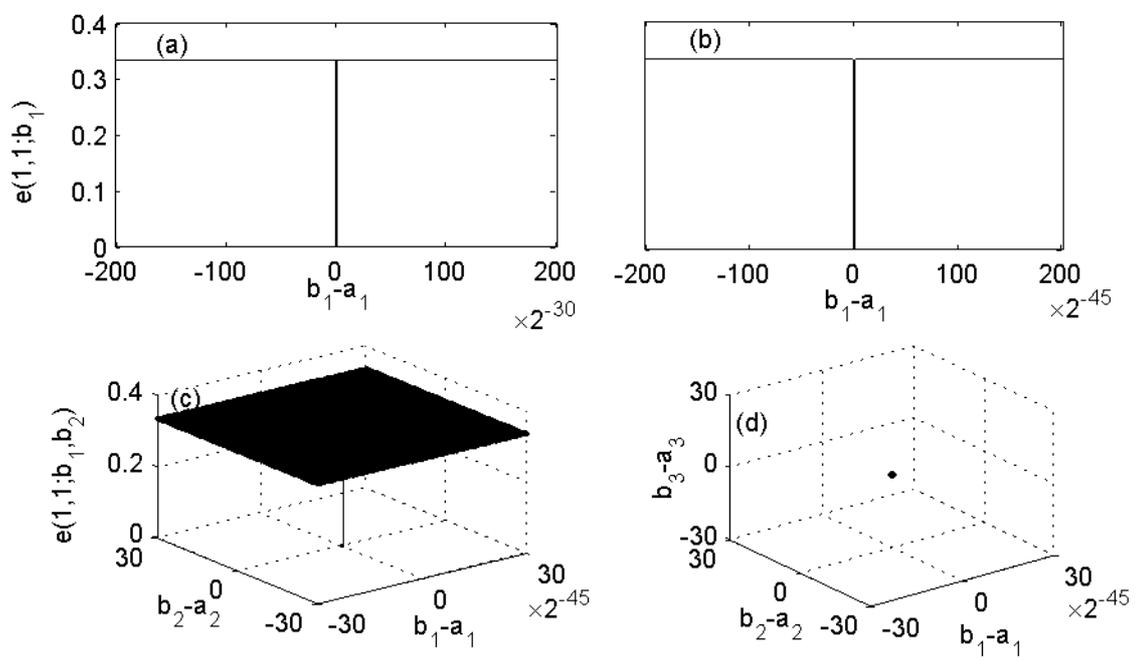
Fig.2. (a), (b) Error function $e(1,1;b_{1})$ defined in Eq.(5) vs the
decryption key parameter $b_{1}$ with different $b_{1}$ detection \
resolutions. $T=10^{8}.$ $b_{2}=b_{3}=a_{2}=a_{3}$. We observe $%
e(1,1;b_{1})=0$ for $b_{1}=a_{1},$ and $e(1,1;b_{1})\simeq \frac{1}{3}$
whenever $b_{1}$ has any mismatch from $a_{1}$ equal to or larger than 2$%
^{-45}.$ (c) $e(1,1;b_{1},b_{2})$ plotted in the $b_{1}-b_{2}$ plan. $%
b_{3}=a_{3}.$ (d) $e(1,1;b_{1},b_{2},b_{3})$ presented in the ($%
b_{1},b_{2},b_{3})$ space. A mesh is plotted black if $e<0.333,$ and left
blank otherwise. With 2$^{-45}$ resolution, only a single black mesh is
observed at $b_{1}=a_{1},$ $b_{2}=a_{2},$ and $b_{3}=a_{3}.$

Fig.3 Scheme of duplex secure speech communication experimental system. User
A and user B are talking over secure telephones. User A talks through A's
Microphone, which produces analog speech signal. A's Analog to Digital
Converter (AD) converts the analog speech into 128Kbit/s digital speech
stream (8K samples a second, 16 bits a sample). A's Source Coder compresses
the digital speech into 8Kbit/s redundancy discarded speech according to a
lossy speech coding standard ITU-T G.729 \cite{20}$.$ (Compression is needed
here so that a 33.6Kbit/s channel can transmit it). A's Encryption unit
encrypts the compressed speech plaintext into ciphertext by using STCC and
AES systems. A's Channel Coder codes the ciphertext into an error correct
code steam \cite{19}. Then a modem modulates the digital code stream into
analog signal, and send the signal to PSTN System. User B receives the
transmitted signal via PSTN and the inverse processing.

Fig.4 The error bit rates $p_{N}$ (squares $\square $, for NNC)$,$ $p_{S}$
(circles $\bigcirc $, for STCC)$,$ and $p_{A}$ (triangles $\triangle $, for
AES)$,$ plotted vs the error bit rate in the noisy transmission channel
noise $p$. $p_{N,S,A}$ are computed by comparing the output signals measured
at gate G2 with the input signal measured at gate G1. All plots are obtained
by averaging ten measurements with each measurement taking 4Mbits. The
vertical bars denote fluctuation ranges. (a) All $p_{N,S,A}$ are measured
without bit error correction. (b) $p_{S,A}$ are measured with bit error
correction while $p_{N}$ not. $p_{S}$ is approximately equal to $p_{N}$ for $%
p\leq \frac{1}{30}$ while $p_{A}>p_{N,S}$ for $p\geq \frac{1}{2000}$.

Fig.5 Analyses of the experimental data of speech signal English word
\textquotedblright welcome\textquotedblright\ said by a female speaker. (a)
Input signal taken from gate G3 of Fig.3. (b)-(g). Received signals taken
from gate G4 of Fig.3. (b), (c) Received signals with NNC and without bit
error correction; (d), (e) with STCC and with bit error correction; (f), (g)
with AES and with bit error correction. $p=\frac{1}{100}$ for (b), (d), (f)
and $p=\frac{1}{30}$ for (c), (e), (g). In each figure the top panel shows
speech signal waveform with horizontal coordinate axis representing time of
0.45 second and vertical axis the amplitude of the waveform (the largest
amplitude of the input signal is normalized to one); the bottom panel
presents pitch (or say, tone of speech) in $f-t$ plane with $f$ being the
pitch frequency and $t$ time when the pitch is taken with the analysis
window of 0.01 second.

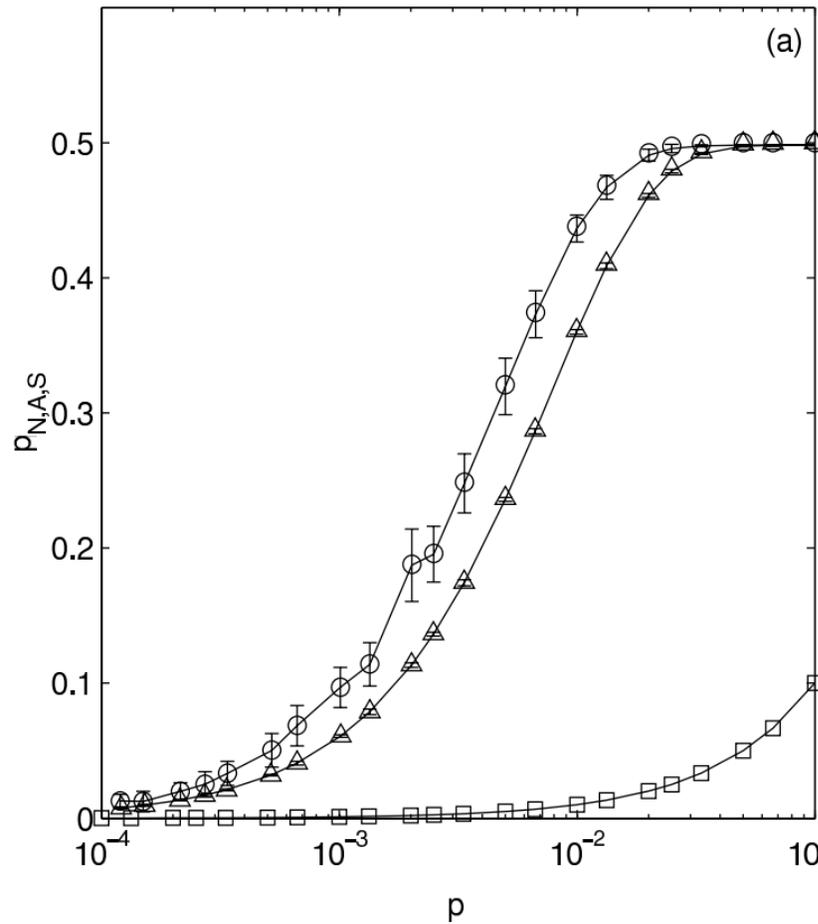
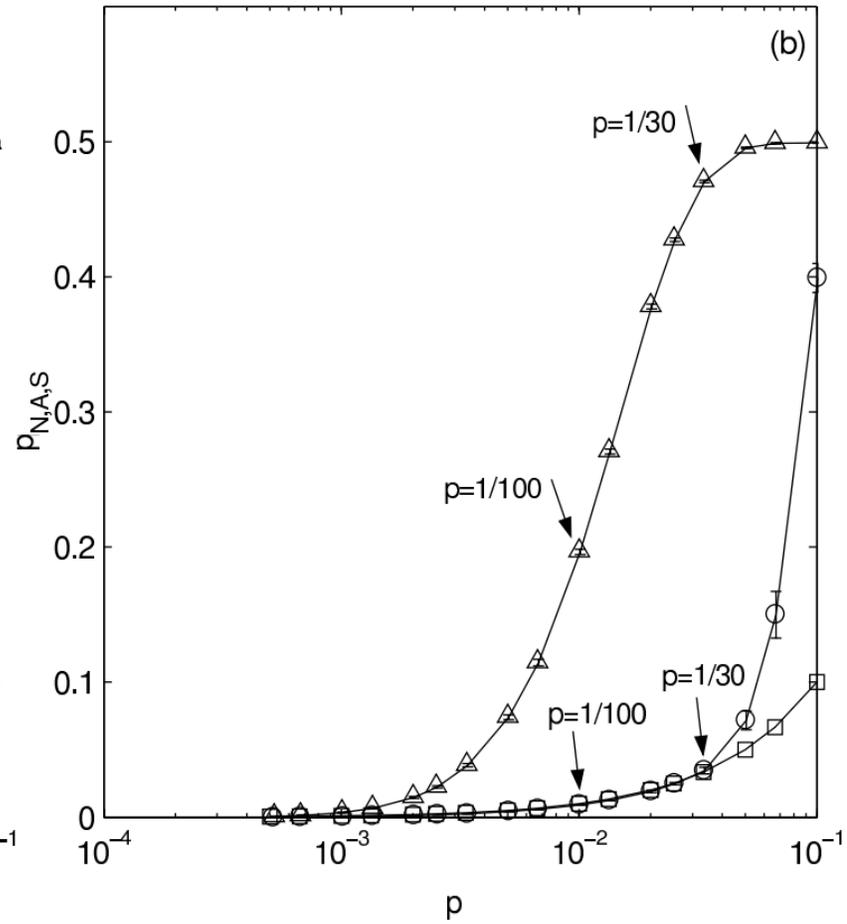
\end{document}